\documentclass[12pt,NJP]{iopart}
\usepackage{graphicx}
\usepackage{color}
\usepackage{bm}
\usepackage[colorlinks,urlcolor=blue,linkcolor=blue,anchorcolor=blue,citecolor=blue]{hyperref}
\usepackage{lineno}

\begin{document}
\title[{QBs in the diatomic $\beta$-FPUT chains}]{$q$-Breathers in the diatomic $\beta$-Fermi-Pasta-Ulam- Tsingou chains}
\author{Lin Deng$^{1}$, Hang Yu$^{1}$, Zhigang Zhu$^{2,1}$, Weicheng Fu$^{3,1}$, Yisen Wang$^{1,*}$, and Liang Huang$^{1,*}$}
\address{$^{1}$ Lanzhou Center for Theoretical Physics, Key Laboratory of Theoretical Physics of Gansu Province, and Key Laboratory of Quantum Theory and Applications of MoE, Lanzhou University, Lanzhou, Gansu 730000, China}
\address{$^{2}$ Department of Physics, Lanzhou University of Technology, Lanzhou, Gansu 730000, China}
\address{$^{3}$ Department of Physics, Tianshui Normal University, Tianshui 741001, Gansu, People’s Republic of China}
\address{$^*$ Authors to whom any correspondence should be addressed.}
\eads{\mailto{wys@lzu.edu.cn}, \mailto{huangl@lzu.edu.cn}}
\vspace{10pt}

\begin{abstract}
$q$-Breathers (QBs) represent a quintessential phenomenon of energy localization, manifesting as stable periodic orbits exponentially localized in normal mode space. Their existence can hinder the thermalization process in nonlinear lattices. In this study, we employ the Newton's method to identify QB solutions in the diatomic Fermi–Pasta–Ulam–Tsingou chains and perform a comprehensive analysis of their linear stability. We derive an analytical expression for the instability thresholds of low-frequency QBs, which converges to the known results of monoatomic chains as the bandgap approaches zero. The expression reveals an inverse square relationship between instability thresholds and system size, as well as a quadratic dependence on the mass difference, both of which have been corroborated through extensive numerical simulations. Our results demonstrate that the presence of a bandgap can markedly enhance QB stability, providing a novel theoretical foundation and practical framework for controlling energy transport between modes in complex lattice systems. These results not only expand the applicability of QBs but also offer significant implications for understanding the thermalization dynamics in complex lattice structures, with wide potential applications in related low-dimensional materials.
\end{abstract}


\maketitle

\section{Introduction}
\label{sec1}

Localized phenomena play a crucial role in the field of condensed matter physics, which relate with a wide range of concepts such as Anderson localization \cite{Lagendijk2009PhysToday,Zhou2023PhysRevLett}, solitons \cite{Zabusky1965PhysRevLett,Kartashov2011RevModPhys,SotoCrespo2016PhysRevLett}, discrete breathers \cite{Ablowitz1973PhysRevLett,Kevrekidis2016PhysRevLett,Duran2023PhysRevE,Chong2024PhysRevResearch} and breather solitons \cite{Velarde2016EPJB,Yu2017NC,Xian2020PhysRevLett}, etc.
Investigating the underlying mechanisms behind these localized states not only deepens our understanding of phenomena such as metal-insulator transitions \cite{Anderson1958PhysRev} and abnormal thermal transport in low-dimensional nonlinear lattices \cite{Xiong2018PhysRevE,Xiong2022PhysRevE}, but also provides theoretical foundations for applications as diverse as managing energy transmission within materials \cite{Duran2023PhysRevE,Chong2024PhysRevResearch} and developing innovative devices for quantum information processing \cite{Mazo2000PhysRevLett}. A key feature of these localized phenomena is the concentration of energy within a limited number of degrees of freedom in the system.

As a typical case, the Fermi–Pasta–Ulam–Tsingou (FPUT) recurrence, first released in a groundbreaking computer experiment conducted by E. Fermi and collaborators, insinuates a localized phenomenon in the normal mode space, where the majority of the energy remains concentrated within the first five low-frequency modes and periodically returns to the originally excited mode \cite{Fermi1955Report, Dauxois2008PhysToday, Gallavotti2007fermi}.
Subsequent investigations revealed that this localization leads to rapid energy distribution across the low-frequency modes up to a cutoff frequency, forming natural packets \cite{Berchialla2004localization}, with the cutoff frequency depending on the magnitude of the initial excitation energy. Much effort has since been dedicated to understanding and explaining the FPUT results, including the role of solitary waves in the Korteweg-de Vries equation \cite{Zabusky1965PhysRevLett}, the stochasticity threshold \cite{Israiljev1965statistical,Chirikov1971}, and the integrability of Toda chains \cite{Toda1967JPS,Benettin2013JSP,Hofstrand2024PRE}.

A deeper understanding of this localization phenomenon requires the concept of $q$-breathers (QBs), proposed as a fundamental type of nonlinear vibrations. QBs are temporally periodic orbits that involve only a few degrees of freedom in the nonlinear lattice, with energy exponentially localized in normal mode space \cite{Flach2005PhysRevLett}.
The original trajectories observed in the FPUT recurrence can be interpreted as perturbations of QBs \cite{Flach2006PhysRevE,FLACH2008PhysicaD}.
As periodic orbits in the nonlinear systems, QBs remain stable in the weakly nonlinear regime, where the energy exchange between QBs and other modes are forbidden. However, in the strongly nonlinear regime, these periodic orbits can be disrupted by various dynamical processes, such as parametric resonance \cite{Yoshimura2000PhysRevE} and Chirikov resonance \cite{CHIRIKOV1979PhysRep}.
A systematic theoretical and numerical analysis of the existence and stability of QBs in the FPUT chains is provided in Ref. \cite{Flach2006PhysRevE}.
In addition, Christodoulidi and colleagues introduced the concept of $q$-tori, which offers exponential local solutions in $q$-space for the FPUT system, effectively bridging QBs and natural packets \cite{Christodoulidi2010PhysRevE, Christodoulidi2013PhysicaD}.

QBs have been extensively verified in various nonlinear monatomic chains, including two-dimensional and three-dimensional $\beta$-FPUT systems \cite{Ivanchenko2006PhysRevLett}, the Bose-Hubbard chain \cite{Nguenang2007PhysRevB, Pinto2009PhysicaD},  and nonlinear Schrödinger lattices \cite{Mishagin2008NJP, Ivanchenko2009JETPLett}, etc. Ivanchenko extended research to disordered FPUT systems, demonstrating that QBs persist as long as the disorder remains minimal, though the instability threshold is highly sensitive to the specific disorder configuration \cite{Ivanchenko2009PhysRevLett}.
Penati and Flach further investigated the tail resonances of QBs, clarifying their role in the pathway toward equipartition \cite{Penati2007Chaos}.
Ivanchenko's subsequent studies on QBs and thermalization in acoustic chains, emphasizing that dynamical localization in normal mode space is predominantly governed by the lowest-order nonlinear terms \cite{Ivanchenko2010JETPLett}. To date, researchers have focused exclusively on monatomic systems, raising the natural question: Do QBs exist in more complex lattices, namely oscillator chains with more than one type atoms per unit cell? If so, how does the bandgap in these lattices affects the behavior of QBs?

The one-dimensional (1D) diatomic chain, as the simplest representative of more complex lattice systems, has garnered significant attention \cite{Dmitriev2009PhysRevB,Vainchtein2016PhysRevE,Fu2019PhysRevE,Pezzi2021threewave,Feng2022JSM,Ngamou2023PhysRevE}.
In this work, we successfully obtain the QB solutions in the diatomic $\beta$-FPUT chains using Newton's Method, and conduct a comprehensive stability analysis. Specifically, in Sec. \ref{sec2}, we introduce the model, providing analytic expressions for the dispersion relation and normal modes, followed by the derivation of the Hamiltonian in term of the normal coordinates.
In Sec. \ref{sec3}, we revisit the theoretical methods for identifying QB solutions, showcasing the energy spectrum of these solutions in the normal mode space and validating these results through numerical simulations.
Section \ref{sec4} focuses on the stability analysis, examining how system size and bandgap influence the stability of QBs. In Sec. \ref{sec5}, we provide a thorough summary and discussion of our findings. These results demonstrate that the system size and bandgap of diatomic chains offer a promising avenue for manipulating the stability of these modes and the intermodal energy exchange channels, which highlights the potential applications and significance of QBs in complex lattice systems.

\begin{figure}[t]
    \centering
    \includegraphics[width=0.7\linewidth]{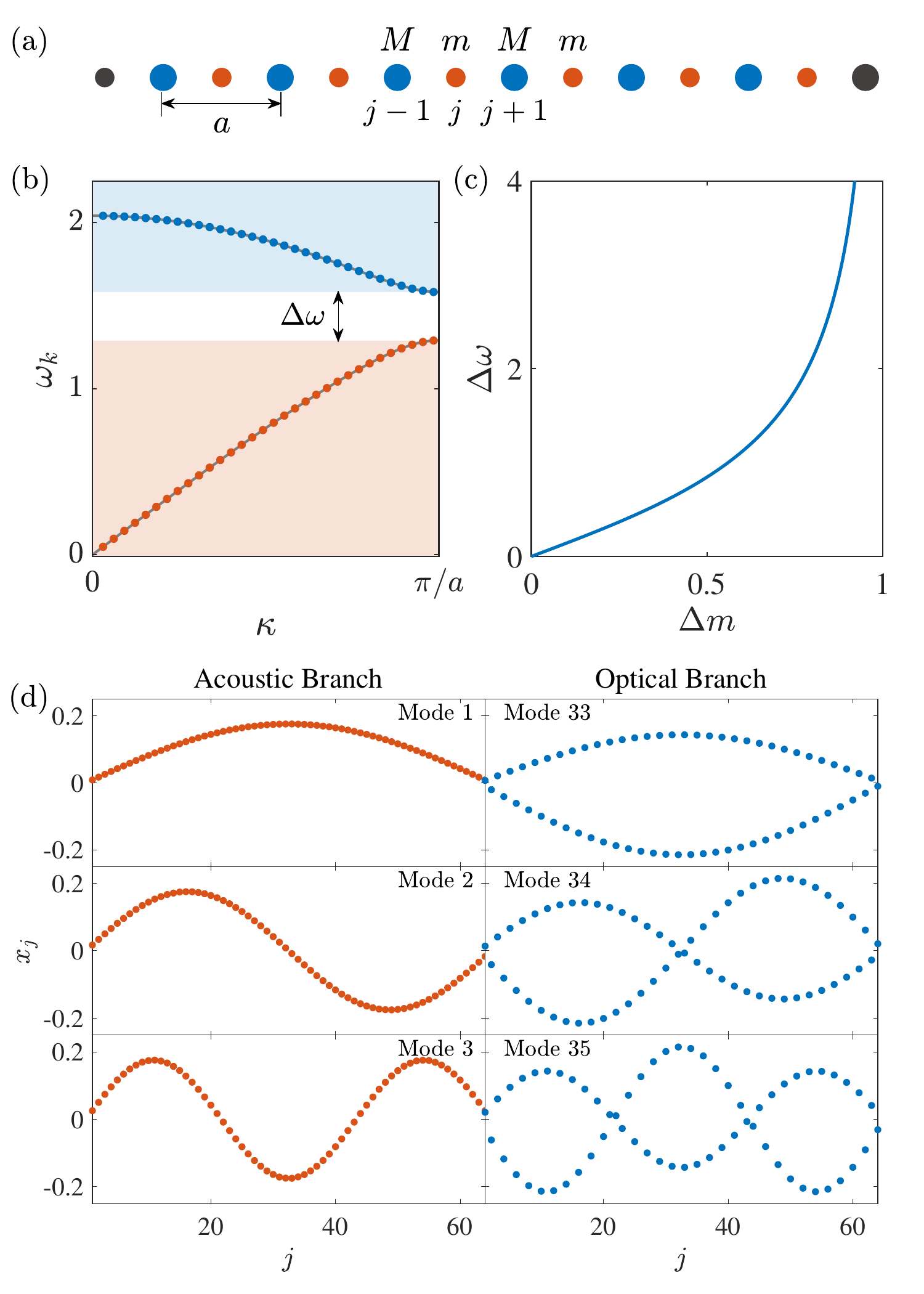}
    \caption{(a) Schematic diagram of 1D diatomic chain with alternating masses. The large blue atoms, with mass $M$, alternate with small red atoms with mass $m$. The black sites are fixed as boundaries.
    (b) Dispersion relation of the diatomic chain with $\Delta m=0.2$ under fixed boundary conditions. The red and blue circles are the eigenfrequencies of the acoustic and optical branches for the system with $N=32$ unit cells. (c) Dependence of the bandgap $\Delta \omega$ on the mass difference ratio $\Delta m$. (d) The morphology of normal modes corresponding to the first three wave vectors.
 }
    \label{fig:model}
\end{figure}

\section{Model}\label{sec2}
We consider a 1D diatomic lattice with alternating masses $M = 1 + \Delta m$ and  $m = 1 - \Delta m$, with $\Delta m \in (0, 1)$ representing the ratio of the mass difference relative to the unit mass. As illustrated in Fig. \ref{fig:model}(a), the system consists of $N$ unit cells, along with two boundary atoms, resulting in a total of $2N+2$ sites. The spacing between neighboring particles is set to unity for simplicity. The Hamiltonian governing the dynamics of this system is given by
\begin{equation}\label{Hamiltonian}
    H =\sum_{j=1}^{2N+1} \left[\frac{p_{j}^2}{2 m_j} +\frac{(x_{j} - x_{j-1})^2}{2} + \frac{\beta (x_{j} - x_{j-1})^4}{4} \right],
\end{equation}
where $x_{j}$ denotes the displacement of atom $j$ from its equilibrium position, $p_{j}$ is the corresponding conjugate momentum, $m_j$ is the mass of particle, $\beta>0$ is the nonlinear parameter.
In this work, fixed boundary condition (FBC) is applied throughout, i.e., $x_{0} = x_{2N+1} = 0$, $p_{0} = p_{2N+1} = 0$.

Using lattice dynamics theory \cite{Born1988dynamical, Dove1993LatticeDynamics, Kittel2005introduction}, the dispersion relation is given by (see \ref{AP_A} for detailed derivation)
\begin{equation}\label{sesan}
    \omega^2_{k,\pm} = {\frac{2}{M m} \left[1 \pm \sqrt{1 - M m \sin^2\left(\frac{\pi k}{2N + 1}\right)} \right]},
\end{equation}
where $k \in \{1, 2, \ldots, N\}$, $\omega_{k,-}$ and $\omega_{k,+}$ denote acoustic and optical branches, respectively.
When $\Delta m \neq 0$, a bandgap $\Delta \omega = \sqrt{2} \left( \frac{1}{\sqrt{1 - \Delta m}} - \frac{1}{\sqrt{1 + \Delta m}} \right)$ naturally exists between the acoustic and optical branches, as sketched in Fig. \ref{fig:model}(b).
Figure \ref{fig:model}(c) shows the dependence of bandgap size \(\Delta \omega\) on the mass difference ratio $\Delta m$.

The normal modes ${\bm v}^{\pm}_k$, in the form of standing waves under FBC, can be obtained by superposing the traveling waves with opposite wave vectors, which can be expressed as
\begin{equation}\label{eigenstates}
v_{j,k}^{\pm}=
\cases{\frac{2}{\sqrt{2N + 1}} e_1^{\pm}(k) \sin\left(\frac{\pi j k}{2N + 1}\right), & mod($j$,2)=1;\\
\frac{2}{\sqrt{2N + 1}}e_2^{\pm}(k)\sin\left(\frac{\pi j k}{2N + 1}\right), & mod($j$,2)=0.\\}
\end{equation}
Here
\begin{equation}
    e_1^{\pm}(k)=\frac{(2 - m \omega_{k,\pm}^2)\sqrt{M} }{\sqrt{(2 - m \omega_{k,\pm}^2)^2M + 4 \cos ^2\left(\frac{\pi k}{2N + 1}\right) m}},
\end{equation}
and
\begin{equation}
    e_2^{\pm}(k)=\frac{2\cos\left(\frac{\pi k}{2N + 1}\right) \sqrt{m}  }{\sqrt{(2 - m \omega_{k,\pm}^2)^2M + 4 \cos ^2\left(\frac{\pi k}{2N + 1}\right) m}},
\end{equation}
are  the components of the polarization vector $\bm{e}^{\pm}(k)$, which describe the relative displacements of the atoms within the same unit cell for a given normal mode.
All these modes form a complete orthonormal basis in analytic form, represented as ${\bm V}=[{\bm v}_1^-, ...,{\bm v}_N^-,{\bm v}_1^+, ...,{\bm v}_N^+]_{2N\times 2N}$, which is convenient for the subsequent theoretical analysis.
The column index of matrix \(\bm{V}\) gives the mode number $q$, with \(q=k\) and \(q=N+k\) corresponding to the acoustic and optical modes, respectively.

The normal coordinates \(Q_q\) are introduced by the canonical transformations
\begin{equation}\label{canonicaltransformation}
     Q_q(t)= \sum_{j=1}^{2N} \sqrt{m_j} x_{j}(t)  {V}_{j,q}.
\end{equation}
In terms of the normal mode coordinates, the Hamiltonian becomes
\begin{eqnarray}
    H = \frac{1}{2} \sum_{q} \left[ P_q^2 + \omega^2_q Q_q^2 \right] + \frac{1}{8}\frac{\beta}{(2N+1)} \sum_{q,l,i,n} C_{q,l,i,n} Q_{q} Q_{l} Q_{i} Q_{n},
    \label{eq:Hamil_normal}
\end{eqnarray}
where $P_q=\dot{Q}_q$, and $C_{q,l,i,n}$ is the nonlinear coefficient that adheres to the selection rule(see \ref{AP_A} for the expressions)
\begin{equation}\label{SelectionRules}
    \pm k_{1} \pm k_{2} \pm k_{3} \pm k_{4} = 0, \pm 2(N+1).
\end{equation}
The harmonic energy of mode $q$ is given by
\begin{equation}\label{modeenergy}
    E_q = \frac{1}{2} (P_q^2 + \omega^2_q Q_q^2).
\end{equation}

\section{$q$-Breather Solutions}
\label{sec3}

\subsection{Newton's Method for Searching QBs}
In systems where the frequency spectrum meets non-resonance conditions $\omega_q \neq n\omega_{q_0}$ for all integers $n$ and $q \neq q_0$, the periodic orbits of a linear system can persist into the regime of nonzero nonlinearity while maintaining a fixed energy level \cite{Flach2006PhysRevE, LYAPUNOV1992, Flach2008AJP}, where these periodic orbits continuously deform as the nonlinearity varies \cite{MacKay1994,Aubry1994concept}. Moreover, periodic orbits corresponding to a specific nonlinearity parameter $\gamma$ can be iteratively obtained using Newton's method, starting from a known solution at a nearby nonlinearity $\gamma'$ as an initial guess, provided that $\vert \gamma - \gamma' \vert $ is sufficiently small. This iterative process provides a quasi-continuous extension from the trivial solution of linear system to the periodic orbit of the system with a given nonlinearity parameter $\gamma$.

Guided by the principle outlined above, QBs can be systematically traced in the diatomic $\beta$-FPUT chains \cite{Marin1996breathers}.
When $\beta = 0$, the Hamiltonian, as given by Eq. (\ref{eq:Hamil_normal}), is quadratic, causing all modes to decouple.
In this scenario, if one mode, referred to as the seed mode $q_0$,  is initially excited, the energy remains localized in this mode, described by $Q_{q} (t) = \delta_{q,q_0}A_{q} \cos(\omega_{q} t+\phi_{q})$, where $A_q$ and $\phi_{q}$ are the amplitude and initial phase of the mode $q$, respectively. Thus, each mode represents a trivial QB solution and can serve as an initial guess for finding QB solutions in systems with a sufficiently small parameter $\Delta\beta$, where $\Delta\beta = 10^{-3}$ is adopted in this work unless otherwise specified.
The solution obtained for a given $\Delta\beta$ can then be used as the initial guess for the QB solution at $2\Delta\beta$. By repeating this iterative process, solutions for $\beta = n\Delta\beta$ can be determined, where $n$ is arbitrary positive integer. In this way, QB solution for specific nonlinearity parameter can be identified from the continuous deformation of the periodic orbits corresponding to the seed mode.

\begin{figure}
        \centering
        \includegraphics[width=0.8\linewidth]{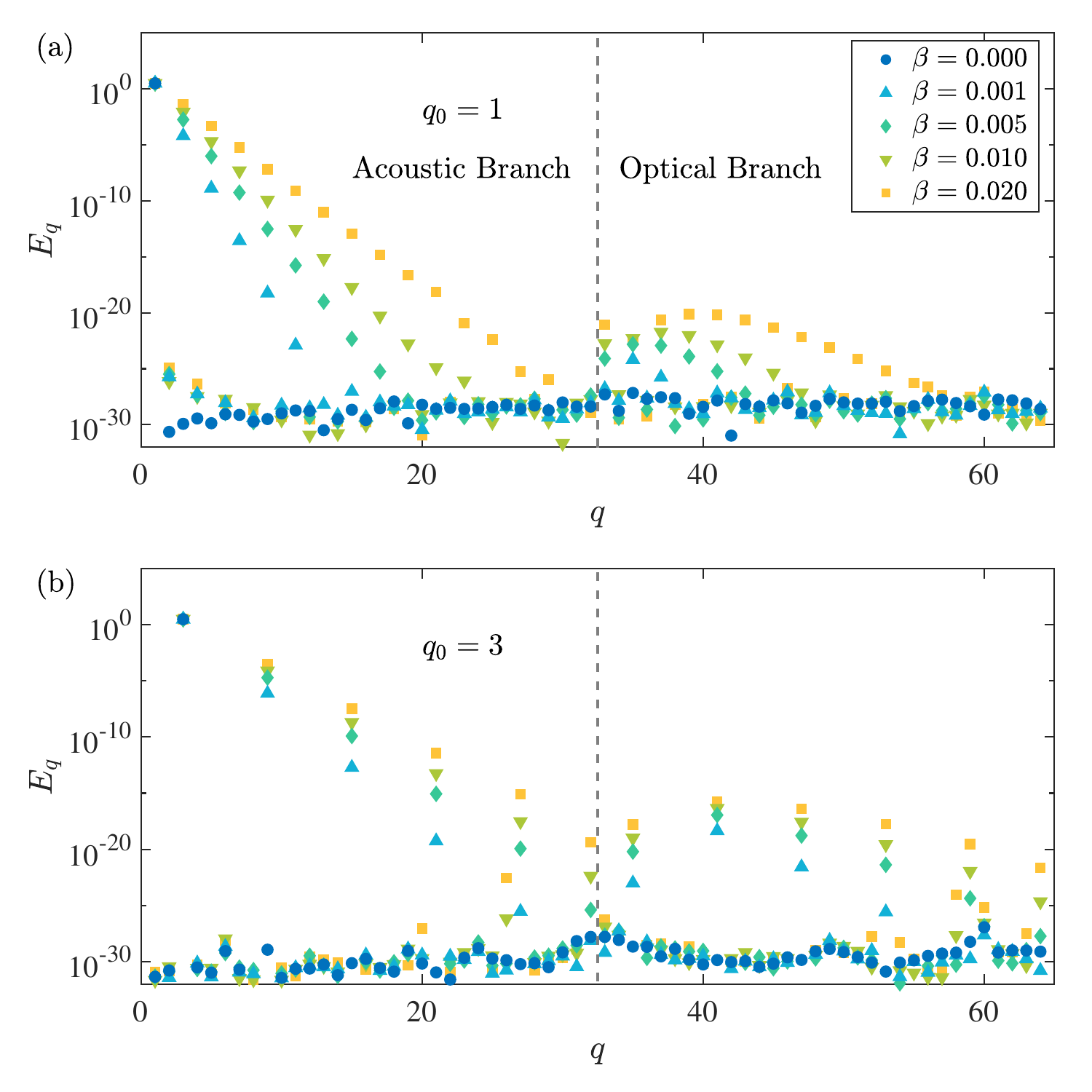}
        \caption{Energy spectrum under the excitation of QBs with varying nonlinearities.
        (a) Seed mode $q_0=1$. (b) Seed mode $q_0=3$. In both panels, the system size and mass difference ratio are $2N = 64$ and $\Delta m=0.5$, respectively. The excitation energy is $E_{\rm total} = 3.16$. The vertical dashed lines indicate the boundary between the acoustic and optical branches.}
        \label{Fig:QBs}
\end{figure}

In this work, QBs are constructed as follows.
We begin by selecting the Poincaré section $S: \{Q_{q_0} = 0, P_{q_0} > 0\}$ in the phase space, and the trajectory of the system intersects this section at the point $ \bm{s}$, represented by a $4N$-dimensional vector $\bm{s} = \{Q_1, Q_2, \ldots, Q_{2N}, P_1, P_2, \ldots, P_{2N}\}$.
The Poincaré map $\bm{I}$ is defined as follows: starting from a point $\bm{s}_0$ on the section $S$, we integrate the system until it returns to this section again, intersecting at point $\bm{s}_1$, which can be expressed as
$\bm{s}_1 = \bm{I}(\bm{s}_0, \beta)$.
The fixed points of the map $\bm{I}$ correspond to the periodic orbits of the system and serve as candidates for QBs.
To apply the implicit function theorem for continuing the fixed point to the case with finite nonlinearity at $\beta \neq 0$, it is necesasry to eliminate the degeneracy caused by energy conservation. To achieve this, we consider a new map in the $(4N-2)$-dimensional phase space
\begin{equation}
\bm{F}:\bm{r}_1 = \bm{F}(\bm{r}_0, \beta),
\label{eq:mapF}
\end{equation}
where $\bm{r} = \{Q_1,Q_2, \ldots, Q_{q_0-1}, Q_{q_0+1}, \ldots, Q_{2N}, P_1,P_2,$ $\ldots, P_{q_0-1}, P_{q_0+1}, \ldots, P_{2N}\}$,
excluding the two variables $Q_{q_0}=0$ and $P_{q_0}$.
Equation (\ref{eq:mapF}) defines a mapping of vector $\bm{r}$ onto itself by fixing $Q_{q_0}=0$, which ensures that the fixed point of the map $\bm{F}$, i.e.,  $\bm{F}(\bm{r}_{\rm f}, \beta) - \bm{r}_{\rm f} = \bm{0}$, is uniquely associated with the fixed point $\bm{s}_{\rm f}$ of the map $\bm{I}$,  representing the same periodic orbit.

In our simulations, the $SBAB_2$ symplectic integrator is utilized to integrate the equations of motion \cite{Skokos2009PhysRevE,Skokos2010PhysRevE}.
We solve for the roots of the equation
\begin{equation}\label{mappingG}
\bm{G}(\bm{r}, \beta) = \bm{F}(\bm{r}, \beta) - \bm{r} = \bm{0},
\end{equation}
by using the Newton's method with the following iteration process
\begin{equation}\label{iterationprocess}
\bm{r}' = \bm{r} - \bm{N}^{-1} [\bm{F}(\bm{r}, \beta) - \bm{r}],
\end{equation} where the Newton matrix $\bm{N}$ for the vector function $\bm{G}(\bm{r})$ is defined as
\begin{equation}\label{Newtonmatrix}
\bm{N}_{i,j} = \frac{\partial G_i (\bm{r}, \beta)}{\partial r_j} = \frac{\partial F_i (\bm{r}, \beta)}{\partial r_j} - \delta_{ij}.
\end{equation}
Here $\partial F_i (r)/\partial r_j$ is the Jacobian matrix of the mapping $\bm{F}$ on the section $S$, and $\delta_{ij}$ is Kronecker delta. Detailed descriptions of these numerical methods can be found in Refs. \cite{Flach2004computational,FLACH2008PhysRep}.
Using the fixed point $\bm{r}_{\rm f}$ and incorporating the corresponding energy adjustment component $P_{q_0} = \sqrt{2E - \sum_{q \neq q_0} P_q^2}$, we ultimately obtain the fixed point $\bm{s}_{\rm f}$. The iteration process is terminated when the condition $\left\| \bm{I}(\bm{s}_{\rm f},\beta)-\bm{s}_{\rm f} \right\| < 10^{-9}$ is satisfied, where $\left\| \bm{s} \right\|=\max \{|\bm{s}|\}$.
This level of precision ensures the accuracy of the periodic orbit solution within the specified tolerance.

\begin{figure}[p]
    \centering
    \includegraphics[width=0.7\linewidth]{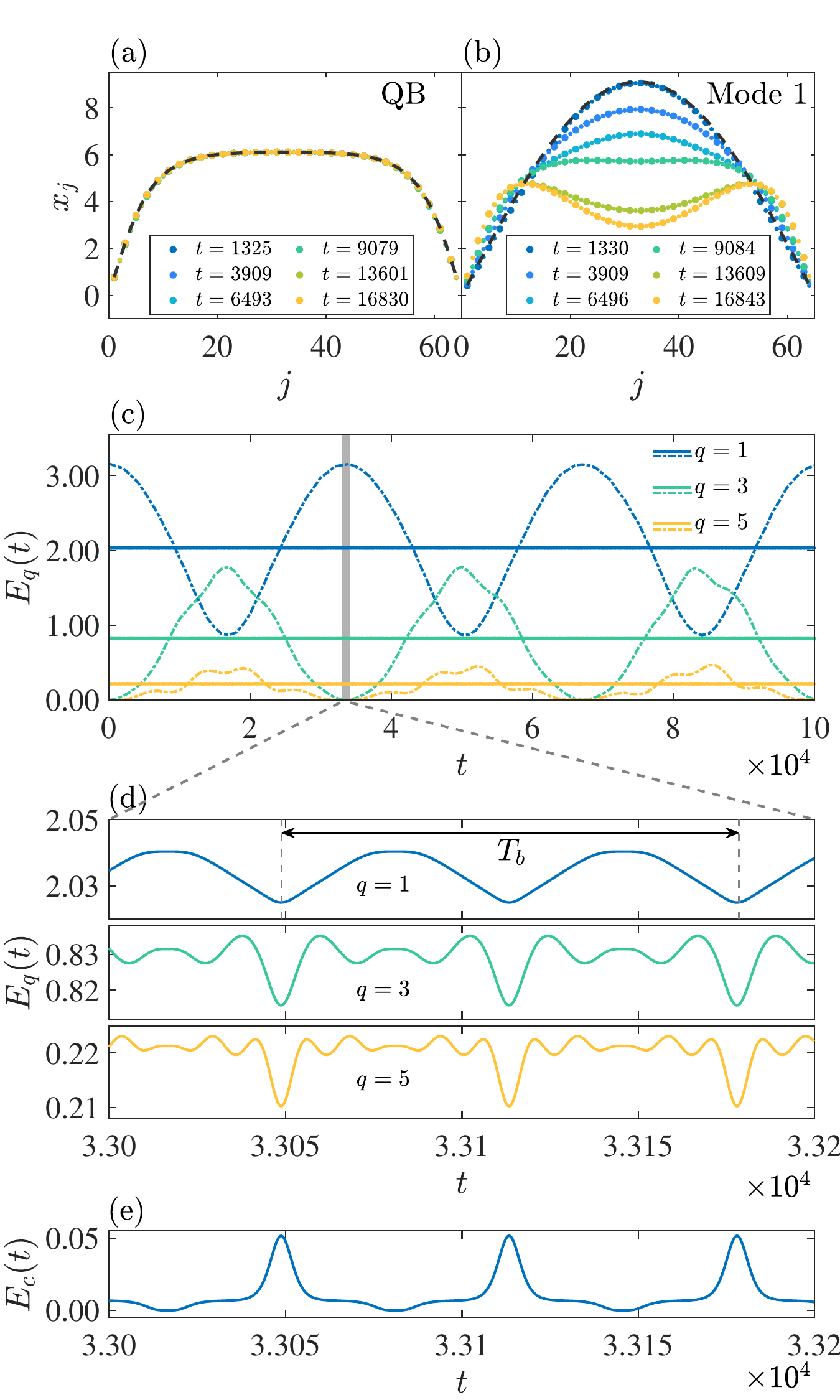}
    \caption{(a, b) Snapshots of the system from MD simulations under the excitation of QB solution (a) and mode $q_0=1$ (b). The black dashed lines present the displacement profile for the QB solution in (a) and mode 1 in (b).
    For both panels, $\beta = 0.08$, $E_{\rm total} = 3.16$, $\Delta m=0.5$, and $2N = 64$.
    (c) Time evolution of energies for the three dominant modes. The solid lines are QB solution, whereas the dashed lines correspond to the excitation with mode 1.
    (d) Zoon-in of the time evolution of QB solution in (c) for the time window marked by the translucent grey plane.
    (e) Time evolution of the total nonlinear energy $E_c(t)$, defined as $E_c(t) = E_{\rm total}(t) - \sum_{q=1}^{2N} E_q(t)$. }
    \label{Fig:compare}
\end{figure}

\subsection{Numerical Results}

As a representative example, we consider a diatomic chain with $2N=$ 64 free particles under FBC, and search for the QB solution.
Figure \ref{Fig:QBs} shows the energy spectrums of the QBs with seed modes $q_0=1$ [Fig. \ref{Fig:QBs}(a)] and $q_0=3$  [Fig. \ref{Fig:QBs}(b)].
In the acoustic branch, the mode energies decrease exponentially as the mode number increases, reflecting an exponential localization of energy in the normal mode space, which is analogous to the QBs observed in the monatomic $\beta$-FPUT chains \cite{Flach2005PhysRevLett}. This type of energy localization can be analytically understood through the modal equation, which incorporates intermodal coupling terms governed by the selection rule [Eq. (\ref{SelectionRules})] \cite{BIVINS1973}.
However, in the optical branch, some modes are also excited, albeit with relatively low energies, indicating weak coupling between acoustic and optical modes due to the mass difference.
Additionally, Figs. \ref{Fig:QBs} (a) and \ref{Fig:QBs}(b) exhibit that as $\beta$ increases, indicative of stronger nonlinearity, the energy distribution of the system becomes more delocalized, with a greater number of normal modes being excited. This behavior aligns with the characteristics of natural packets \cite{Berchialla2004localization}.

To elucidate the correlation between QB and thermalization dynamics of the system, we conduct MD simulations under different initial conditions.
Figure \ref{Fig:compare}(a) displays the snapshots of the system under the excitation of the QB solution with seed mode $q_0=1$ at $\beta=0.08$, showing a flatter peak compared to mode 1, as seen in the first panel of Fig. \ref{fig:model}(d).
A notable observation is that, all curves closely match the displacement profile of the QB solution when $P_{q_0}=0$, and the displacement profiles remain consistent across different cycles throughout the simulations, indicating a low-dimensional torus in phase space with minimal energy exchange between normal modes, also shown in the Supplementary Video.
Correspondingly, the time evolution of the linear energies $E_q(t)$ for the three dominant modes involved in the QB remains almost constant, exhibited by the basically straight lines in Fig. \ref{Fig:compare}(c). This means that the QB solution represents a dynamical structure consisting of a packet of normal modes satisfying the selection rule, effectively preventing thermalization by confining nearly all the system's energy.
In contrast, when the system is excited by the first normal mode,  the displacement profiles of the system evolve progressively [Fig. \ref{Fig:compare}(b)], indicating the excitation of additional modes [Fig. \ref{Fig:compare}(c)]. In fact, the system experiences FPUT recurrence on a time scale of approximate $3.3\times 10^4$, as depicted by the dashed lines in Fig. \ref{Fig:compare}(c).

Another key feature of QBs is their periodicity, which is evident from the constant energy maintained within each mode. Under these conditions, each mode undergoes regular, periodic vibrations. The periodicity of the QB is further supported by the minor fluctuations in mode energy, with the magnitude of these fluctuations being approximately 0.6\% of the total energy [Figs. \ref{Fig:compare}(d) and \ref{Fig:compare}(e)].
Figure \ref{Fig:compare}(d) provides a close-up view of Fig. \ref{Fig:compare}(c), and all modes exhibit oscillations with the same period $T_{b} = 2\pi/\hat{\omega}_b \simeq 129.32$, where $T_b$ and $\hat{\omega}_b$ are the period and frequency of QB, respectively.
This uniform periodicity across all modes is a hallmark of QB, reinforcing the notion that QBs act as robust, stable dynamical structures in nonlinear systems.
The QB period $T_b$ is slightly smaller than that of the first normal mode, i.e., $T_{q_0} = 2\pi/\omega_{q_0} \approx 130.01$, which is due to the frequency shift induced by the nonlinearity.

\section{Stability Analysis of QBs}
\label{sec4}

\subsection{Stability Estimation Based on the Floquet Theory}
The stability analysis of QBs is essential for understanding their function as fundamental periodic orbits in normal mode space, especially in relation to the thermalization process in nonlinear lattices. It allows for predictions regarding the long-term behavior of QBs, determining whether these periodic orbits will persist or decay over time. Stable QBs indicate sustained localized energy states, which can inhibit or significantly delay the thermalization process, maintaining the system in a non-equilibrium state. In contrast, unstable QBs result in the rapid redistribution of energy among normal modes, thereby accelerating the system's path toward thermal equilibrium. Furthermore, understanding the stability of QBs provides valuable insights into the energy distribution and transfer among modes, underscoring the significant role of localized energy states in regulating energy flow within nonlinear systems.

\begin{figure}[t]
    \centering
    \includegraphics[width=0.7\linewidth]{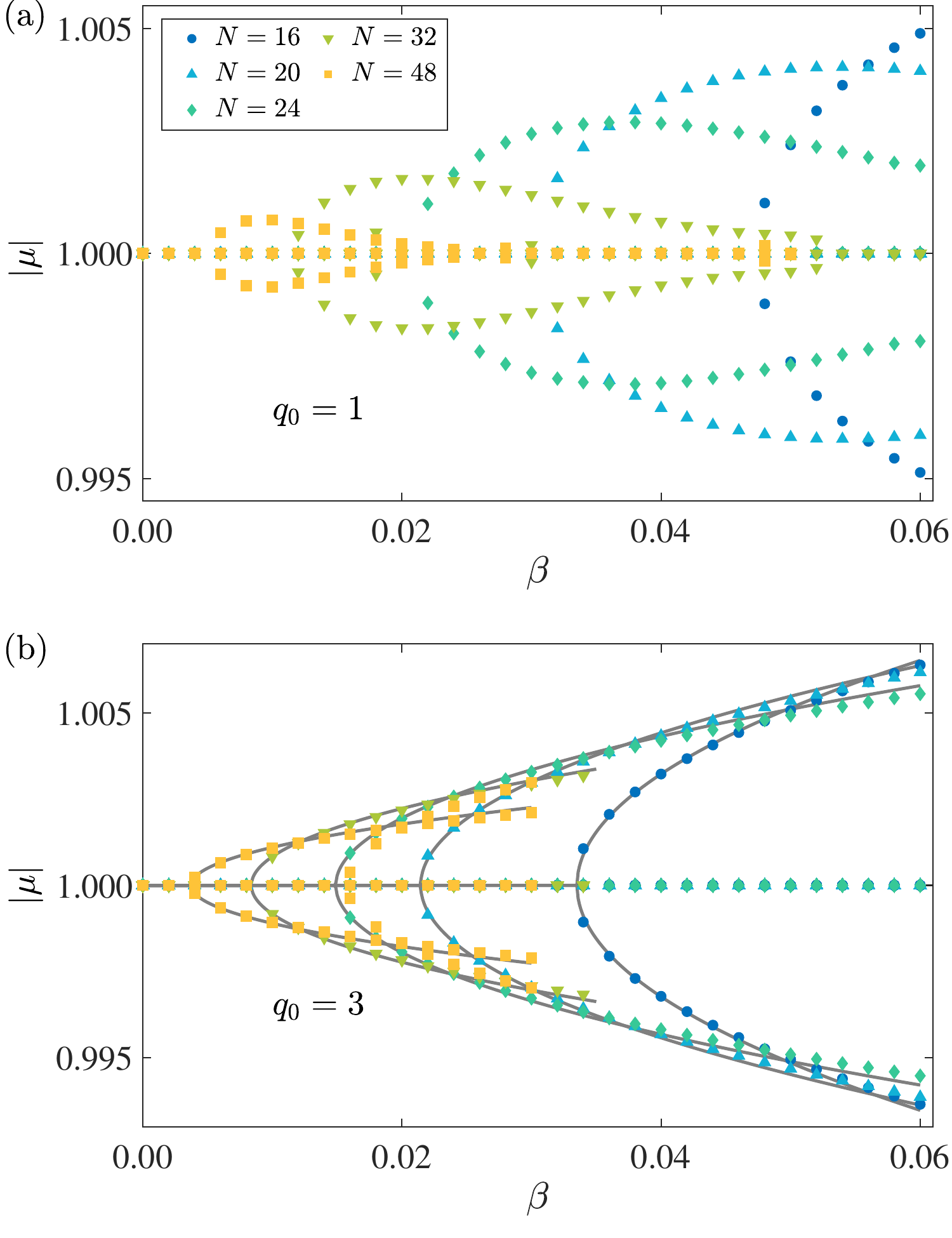}
    \caption{Floquet multipliers $\vert\mu\vert$ for QBs as a function of the nonlinear parameter $\beta$ for systems with different sizes $N$. (a) The seed mode $q_0=1$. (b) The seed mode $q_0=3$. These two panels share the same legend. The grey solid lines in panel (b) represent the theoretical results given by Eq. (\ref{Floquetmultiplier}). In both panels, the simulation parameters are \(E_{\rm total} = 3.16\),  $\Delta m=0.5$, and $N = 32$.
 }
    \label{fig4}
\end{figure}

To ascertain the stability of QBs in the diatomic $\beta$-FPUT model, we perform a linear stability analysis following the approach outlined in the literature \cite{Flach2005PhysRevLett, Flach2006PhysRevE, Flach2004computational, FLACH1998PhyRep, Aubry1997PhysicaD, Campbell2004PhysToday}.
The equation of motion for mode $q$ can be written as
\begin{equation}\label{EofM}
\ddot{Q}_q + \omega_q^2 Q_q = -\rho \sum_{l,i,n=1}^{2N} C_{q,l,i,n} Q_l Q_i Q_n,
\end{equation}
where $\rho = \beta/[2(2N+1)]$.
We assume that the expression of QB solution takes the form
\begin{equation}
\label{QB}
\hat{Q}_q(t) = \delta_{qq_0} A_q \cos(\hat{\omega}_b t) + O(\rho^2).
\end{equation}
By applying standard secular perturbation techniques, we approximate the QB frequency $ \hat{\omega}_b$ as (see \ref{AP_B} for detailed derivation) \cite{Nayfeh1995NonlinearOscillations}
\begin{equation}\label{QBfrequency}
\hat{\omega}_b = \omega_{q_0} \left[ 1 + \frac{9\beta E_{q_0}}{8(2N+1)}  + O(\rho^2) \right],
\end{equation}
where $E_{q_0} = \omega_{q_0}^2 A_{q_0}^2/2$ is the energy localized in the seed mode.
Considering an infinitesimal perturbation $\xi_q$ to the QB solution $\hat{Q}_q(t)$, such that $Q_q(t) = \hat{Q}_q(t) + \xi_q$, the equation governing the time evolution of $\xi_q$ becomes
\begin{equation}\label{perturbedequation}
\ddot{\xi}_q + \omega_q^2 \xi_q = -3\rho \sum_{l,i,n=1}^{2N} C_{q,l,i,n} \hat{Q}_l \hat{Q}_i \xi_n + O(\rho^2, \xi_n).
\end{equation}
Substituting the expression of $\hat{Q}_{q_0}$ [Eq. (\ref{QB})] into  Eq. (\ref{perturbedequation}), the equations of motion for all the modes can be reformulated as Mathieu equation in the matrix form:
\begin{equation}\label{Mathieu}
\ddot{\bm{\xi}} + \bm{\Omega}^2\bm{\xi} + h[1+\cos(2\hat{\omega}_b t)] \bm{B}\bm{\xi} = O(\rho^2, \bm{\xi}),
\end{equation}
where $\bm{\xi} = [\xi_1, ..., \xi_{2N}]^T$, $\bm{\Omega}^2 = {\rm diag}[\omega_1^2,...,\omega_{2N}^2]$,  $h = 3\rho E_{q_0}$ is a small parameter, and the coupling matrix $\bm{B} = [b_{qn}]_{2N\times 2N}$ is defined with $b_{qn}\equiv C_{q,q_0,q_0,n}/\omega_{q_0}^2$.
This reformulation allows us to analyze the stability of QBs and how energy is exchanged between modes.

The nonlinearity strength in the $\beta$-FPUT lattice is generally characterized by the product $\beta\varepsilon$, where $\varepsilon=E_{\rm total}/(2N)$ represents the specific energy.  Since either $\beta$ or $\varepsilon$ can be rescaled to any positive value through a variable transformation \cite{Karve2024Chaos}, one of these parameters can be held constant to study the relationship between nonlinearity and the other.
In this study, we fix the specific energy $\varepsilon = 1.58/32\simeq 0.049$, as used in the original FPUT experiments,
and investigate how the stability of QB solutions depends on the anharmonic parameter $\beta$.

By treating $h$ and $\hat{\omega}_b$ as independent parameters, we analyze the parametric resonance of Eq. (\ref{Mathieu}) and obtain an estimate of the Floquet multiplier that incorporates the effects of mass difference ratio $\Delta m$
\begin{equation}
\label{Floquetmultiplier}
    |\mu_{\pm}| = 1 \pm \frac{\pi^3(1+3\Delta m^2)}{4(2N+1)^2} {\rm Re} \left[ \sqrt{R-1+O\left(N^{-2}\right)}\right],
\end{equation}
where $R = {6\beta E_{q_0} (2N+1)}/{[\pi^2 (1 + 3\Delta m^2)]}$.
Assuming $E_{q_0} \gg \sum_q E_{q \neq q_0}$ indicated by Eq. (\ref{QB}), we approximate $E_{q_0} \approx E_{\rm total} = 2N\varepsilon$. Substituting this, we get (see \ref{AP_C} for detailed derivation)
\begin{equation}
R \approx \frac{6\beta\varepsilon (2N)^2}{\pi^2(1+3\Delta m^2)}.
\end{equation}
A bifurcation occurs when $R_c = 1 + O\left(1/N^2\right)$, which gives the instability threshold for the QB orbit as
\begin{equation}
\label{beracr}
    \beta_{\rm c} = \frac{\pi^2}{24N^2\varepsilon}(1+3\Delta m^2).
\end{equation}
Below this critical value \(\beta_{\rm c}\), QBs are stable.
{It is obvious that as the mass difference approaches zero, Eq. (\ref{beracr}) naturally reduces to the result for monoatomic chains \cite{Flach2006PhysRevE}.}
As \(N\) approaches infinity, the threshold \(\beta_{\rm c}\) trends toward zero, indicating that QBs can not exist in the thermodynamic limit.

\subsection{Numerical Verification}
To analyze the stability of QBs, we linearize the phase space flow around $\hat{Q}_q(t)$ and numerically integrate this flow over one period of the QB. We then compute the symplectic Floquet matrix and diagonalize it to evaluate stability. A QB orbit is considered stable if all the eigenvalues $\mu$ of the Floquet matrix have absolute values equal to 1; otherwise, the orbit is deemed unstable \cite{Flach2005PhysRevLett}.
Figure \ref{fig4} illustrates the eigenvalues $\vert\mu\vert$ as a function of the nonlinearity parameter $\beta$ for different system sizes $N$.
For weak nonlinearity, all eigenvalues remain at $\vert \mu\vert=1$, and the QBs are stable. However, as $\beta$ increases, a bifurcation arises in the diagram, marking the instability threshold $\beta_c$, as illustrated in Figs. \ref{fig4}(a) and \ref{fig4}(b). This bifurcation leads to the instability of the QB's periodic orbit.
Notably, for $q_0=1$ in Fig. \ref{fig4}(a), the relationship between $\vert \mu \vert$ and $\beta$ is non-monotonic. For example, the bifurcation diagrams for $N=32$ and $N=48$ display water drop-shaped profiles.
As $\beta$ increases, QBs experience a sequence of transitions: initially stable, becoming unstable at intermediate nonlinearity, and restabilizing at higher values of $\beta$. This behavior is akin to the instability islands phenomenon observed in extended FPUT chains with long-range interaction, where instabilities emerge in narrow intervals of excitation energy when $\beta$ is fixed \cite{Miloshevich2015PREInstabilities}.

\begin{figure}[t]
    \centering
    \includegraphics[width=0.8\linewidth]{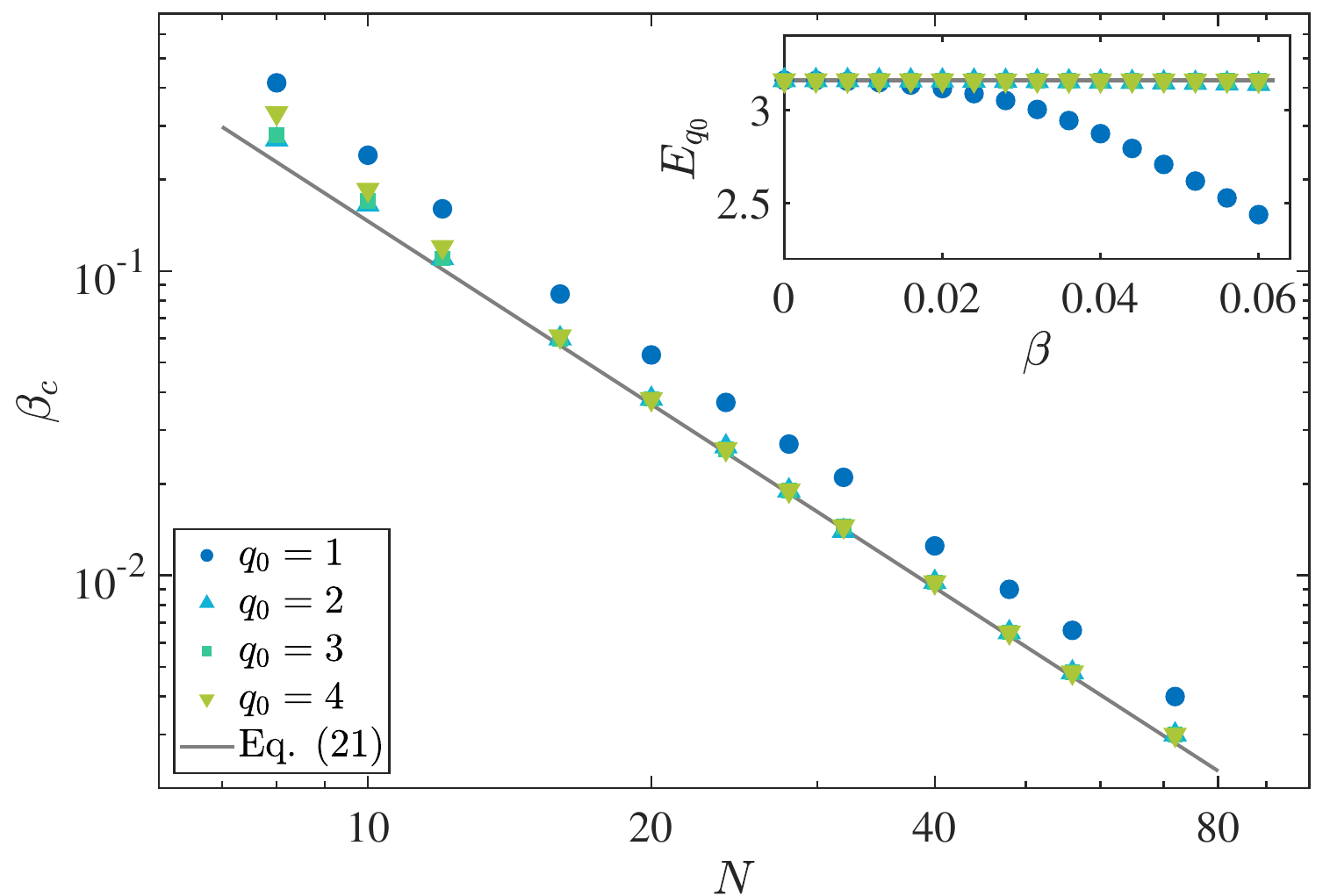}
    \caption{Dependence of the stability threshold $\beta_{c}$ on system size $N$. The symbols represent results from numerical simulations, while the grey solid line denotes the theoretical prediction from Eq. (\ref{beracr}). Note that $\Delta \beta=10^{-4}$ is used for $N>40$ in the iterative process.  The inset illustrates the dependence of $E_{q_0}$ on $\beta$ with $2N=64$ for different seed modes, where the solid line indicates the total energy $E_{\rm total}$.
}
    \label{fig:size}
\end{figure}

An obvious feature revealed by Eq. (\ref{beracr}) is that the instability threshold is inversely proportional to the square of the system size. As shown in Figs. \ref{fig4}(a) and \ref{fig4}(b), larger systems have smaller stability thresholds $\beta_{c}$, consistent with the findings in 1D monoatomic chains \cite{Flach2005PhysRevLett}.
To further illustrate this, Fig. \ref{fig:size} presents the dependence of the stability threshold \(\beta_{c}\) on system size plotted on a log-log scale. It is shown that the data for all QBs with varying seed modes follow an excellent inverse square relationship.
This relationship underscores the fact that as system size increases, the QB orbits become increasingly susceptible to instability, reflecting the challenge of maintaining localized energy states in larger nonlinear systems.
For QBs with $q_0=1$, however, there is a noticeable discrepancy between the numerical results (solid blue circles) and the theoretical
prediction (solid grey line) given by Eq. (\ref{beracr}).
For QBs with other seed modes, the numerical data closely match the theoretical prediction, with only slight deviations observed when the system size is small.

\begin{figure}
    \centering
    \includegraphics[width=0.7\linewidth]{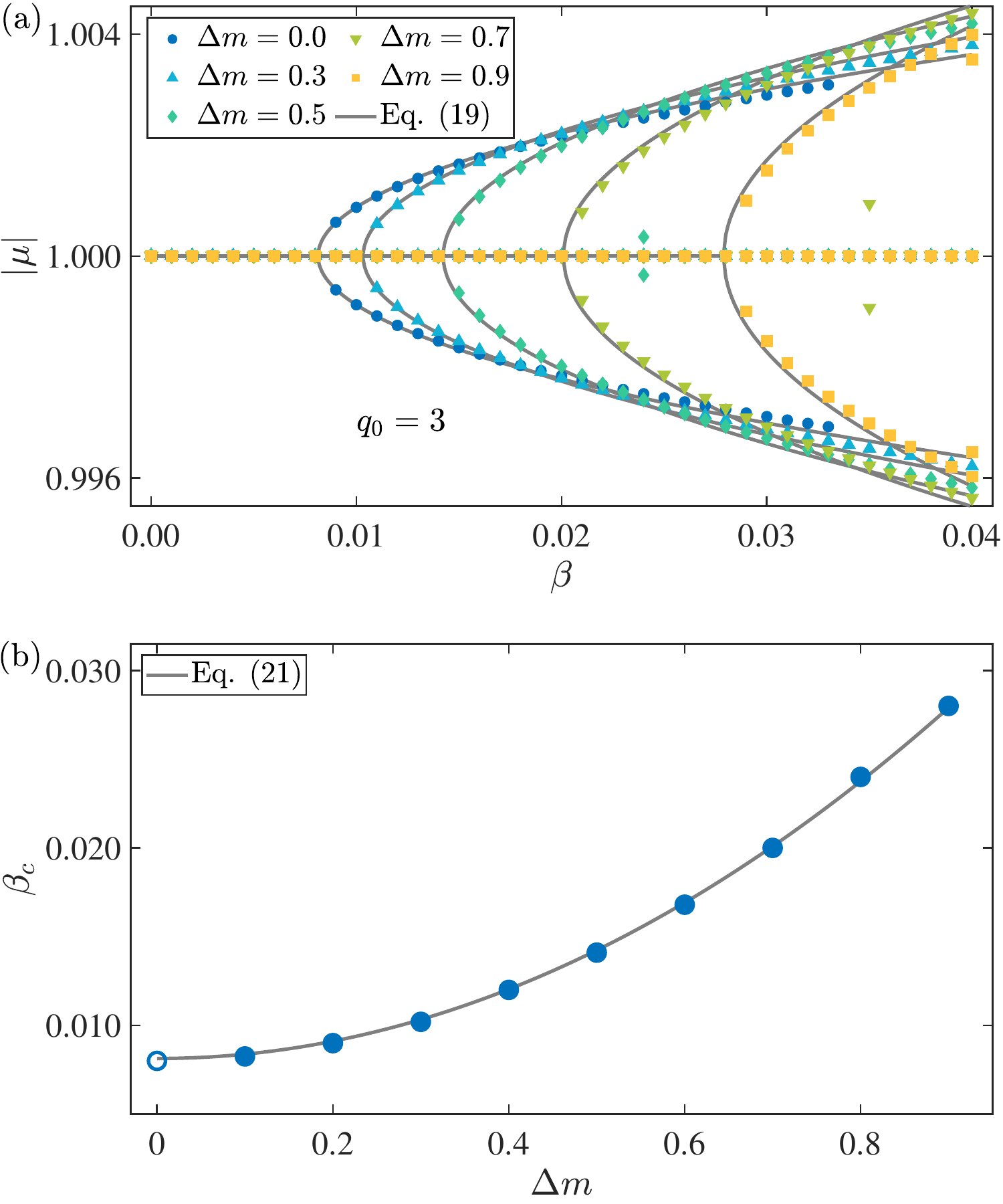}
    \caption{(a) Floquet multipliers as a function of the nonlinearity parameter \(\beta\)  under different  mass difference ratios $\Delta m$ with $2N=64$. (b) The stability threshold \(\beta_{c}\) as a function of \(\Delta m\). The grey solid lines represent the theoretical prediction based on Eq. (\ref{Floquetmultiplier}) in (a) and Eq. (\ref{beracr}) in (b).
}
    \label{fig:deltaM}
\end{figure}

The discrepancy for $q_0=1$ stems from the limitations of Eq. (\ref{QB}), which assumes that nearly all energy is concentrated in the seed mode, i.e., $E_{q_0}\simeq E_{\rm total}$.
The inset of Fig. \ref{fig:size} displays the dependence of $E_{q_0}$ on the nonlinearity parameter $\beta$ for the QB solutions, with the total energy remaining constant, as indicated by the solid grey line.
For $q_0 = 2, 3, 4$, the values of $E_{q_0}$ align with the reference line representing $E_{\rm total}$, contributing to the good agreement between theoretical predictions and numerical simulations.
In contrast, for $q_0 = 1$, $E_{q_0}$ is considerably lower than $E_{\rm total}$, especially in the regime of strong nonlinearity, leading to higher instability thresholds than the theoretical predictions. Thus the theoretical curves are not shown Fig. \ref{fig4}(a).

The diatomic chain is characterized by a bandgap $\Delta \omega$, which is determined by the mass difference ratio $\Delta m$.
Equation (\ref{beracr}) suggests a quadratic relationship between stability threshold $\beta_{c}$ and $\Delta m$.
Figure \ref{fig:deltaM}(a) showcases the bifurcation diagrams of the QB solutions for the systems with varying mass difference ratios $\Delta m$.
It is evident that the monatomic chain, where $\Delta m=0.0$, has the smallest bifurcation point as marked by the empty circles. As $\Delta m$ increases, the bifurcation point progressively shifts to larger values, indicating that the instability threshold $\beta_{c}$ increases with larger mass difference.
Furthermore, Fig. \ref{fig:deltaM}(b) empirically confirms the parabolic dependence of $\beta_{c}$ on the mass difference, validating the theoretical prediction that $\beta_c$ increases with the square of the mass difference. This observation underscores the crucial role of mass difference in determining the stability characteristics of QBs in the diatomic chain.

\section{Summary and Discussions}\label{sec5}

In this work, we successfully obtained the QB solutions with Newton's Method and conducted a comprehensive analysis on its stability in diatomic $\beta$-FPUT chains.
Through a combination of analytical analysis and numerical simulations, we demonstrate that QBs remain stable at weak nonlinearity but undergo bifurcations leading to instability as $\beta$ increases. Specifically, we observed non-monotonic behavior in the bifurcations diagram, a phenomenon reminiscent of instability islands in long-range FPUT chains \cite{Miloshevich2015PREInstabilities}.
Furthermore, we examined the influence of i) system size, and ii) mass difference on the stability of QB solutions.
i) The stability threshold decreases with increasing system size, following an inverse square relationship. However, deviations were observed for QB solutions when the first mode was used as the seed mode, as the assumption that all the energy is concentrated in the seed mode breaks down.
ii) A key result of our analysis is the quadratic dependence of the stability threshold on the mass difference. Larger mass differences correspond to higher instability thresholds, indicating that mass heterogeneity can enhance the stability of localized modes in diatomic chains.
These findings extend the applicability of QBs, providing deeper insight into energy localization and thermalization in nonlinear systems.

Diatomic chains, with their inherent bandgap, offer a unique advantage over monatomic systems in controlling wave propagation, making them particularly useful in materials with tailored thermal and mechanical properties.
Our results underscore the critical influence of system size and mass difference on the stability of localized modes, which not only deepens the understanding of energy localization and thermalization in more intricate nonlinear systems but also presents broader implications for developing advanced materials that require precise energy flow management.
By manipulating the stability threshold, it becomes possible to achieve finer control over energy localization, with potential applications in designing nonlinear metamaterials, thermal insulators, and waveguides.

\section*{Data availability statement}
All data that support the findings of this study are included within the article.

\section*{Acknowledgments}
We thank Prof. Sergej Flach for illuminating discussions. 
This work was supported by NSFC under Grant Nos. 11905087, 12175090, 11775101, 12247101, 12465010 and 12247106, by the 111 Project under Grant No. B20063, and by NSF of Gansu Province under Grant No. 20JR5RA233. W. Fu also acknowledge support by the Youth Talent (Team) Project of Gansu Province; the Innovation Fund from Department of Education of Gansu Province (Grant No.~2023A-106).

\section*{Conflict of interest}
The authors declare that they have no known competing financial interests or personal relationships that
could have appeared to influence the work reported in this paper.

\appendix
\section{Intrinsic Frequencies and Normal Modes}
\label{AP_A}

This section presents a detailed derivation of the Hamiltonian in the normal mode space.

According to the lattice dynamics theory, the dispersion relation for 1D diatomic lattice with periodic boundary condition is expressed as \cite{Born1988dynamical, Dove1993LatticeDynamics, Kittel2005introduction}
\begin{equation}\label{A2}
\omega^{2}_{v|\bm{\kappa}} = {\frac{M + m}{M m} \left[1 \pm \sqrt{1 - \frac{4 M m}{(M + m)^2} \sin^2\left(\frac{\bm{\kappa} \cdot \bm{a}}{2}\right)} \right]},
\end{equation}
and the corresponding eigenstates in the form of traveling waves are
\begin{equation}\label{A6}
{V}_{l|j,v|\bm{\kappa}} =\frac{1}{\sqrt{N}} e_j(v|\bm{\kappa}) e^{{\rm i} \bm{\kappa} \cdot \bm{r}_0(l|j)},
\end{equation}
where $v$ indexes the branches in the dispersion, $\bm{\kappa}$ is the wave vector, and $\bm{r}_0(l|j)$ denotes the equilibrium positions of atom \(j\) in unit cell \(l\).

For systems with FBC, the normal modes are constructed by superposing traveling wave eigenstates with opposite wave vectors, leading to the standing wave solution
\begin{equation}\label{A8}
{V}_{l|j,v|\bm{\kappa}} = \frac{\rm i}{\sqrt{2N + 1}} \left[e_j(v|\bm{\kappa}) e^{-{\rm i} \bm{\kappa} \cdot \bm{r_0}(l|j)} - c.c.\right],
\end{equation}
or equivalently:
\begin{equation}\label{A9}
{V}_{l|j,v|\bm{\kappa}} = \frac{2}{\sqrt{2N + 1}} e_j(v|\bm{\kappa}) \sin(\bm{\kappa} \cdot \bm{r_0}(l|j)),
\end{equation}
where the wave vector $\bm{\kappa}$ can only take discrete values under FBC
\begin{equation}\label{A7}
\bm{\kappa} = \frac{\pi k}{2N+1}\bm{e}_x ,\,  k \in \{ 1,2, \ldots, N\}.
\end{equation}
These eigenstates are orthogonal and normalized, as expressed by
\begin{equation}\label{A10}
\sum_{l=1}^{N} \sum_{j=1}^{2}{V}_{l|j,v_1|\bm{\kappa}_1} {V}_{l|j,v_2|\bm{\kappa}_2} = \delta_{v_1,v_2} \delta_{\bm{\kappa}_1,\bm{\kappa}_2},
\end{equation}
where $\delta$ is Kronecker delta.

The Hamiltonian of the system in the normal mode representation is given by
\begin{equation}\label{A11}
H=H_2+H_4,
\end{equation}
where
\begin{equation}\label{A12}
\eqalign{ H_2 &= \frac{1}{2} \sum_{v|\bm{\kappa}} \left[ P_{v|\bm{\kappa}}^2+ \omega^2_{v|\bm{\kappa}} Q_{v|\bm{\kappa}}^2 \right],\cr
H_4 &= \frac{1}{8}\frac{\beta}{(2N+1)} \sum_{q,l,i,n} C_{q,l,i,n} Q_{q} Q_{l} Q_{i} Q_{n}.}
\end{equation}
Here the interaction coefficient matrix \(C_{q,l,i,n}\) is
\begin{equation}\label{A13}
\eqalign{
C_{q,l,i,n} & =  C_{v_1|\bm{\kappa}_1,v_2|\bm{\kappa}_2,v_3|\bm{\kappa}_3,v_4|\bm{\kappa}_4} \cr
  & =\sum_{\pm}(-1)^c \delta_{\pm \bm{\kappa}_1 \pm \bm{\kappa}_2 \pm \bm{\kappa}_3 \pm \bm{\kappa}_4, 0}   \cr
  &\times \bigg\{ \prod_{s=1}^{4}\left[ \frac{e_2(v_s|\bm{\kappa}_s)}{\sqrt{m}} e^{\pm {\rm i}  \frac{\bm{\kappa}_s \cdot \bm{a}}{2}} - \frac{e_1(v_s|\bm{\kappa}_s)}{\sqrt{M}} \right] \cr
  &+\prod_{s=1}^{4}\left[ \frac{e_1(v_s|\bm{\kappa}_s)}{\sqrt{M}} e^{\pm {\rm i}  \frac{\bm{\kappa}_s \cdot \bm{a}}{2}} - \frac{e_2(v_s|\bm{\kappa}_s)}{\sqrt{m}} \right]\bigg\}.
}
\end{equation}
where $c$ represents the number of negative signs.

\section{Frequency of QBs}
\label{AP_B}

This section briefly derives the QB frequency $\hat{\omega}_b$ based on the standard secular perturbation techniques \cite{Nayfeh1995NonlinearOscillations}.
Substituting Eq. (\ref{QB}) into Eq. (\ref{EofM}) yields
\begin{equation}\label{A14}
-(\hat{\omega}_b^2-\omega_{q_0}^2) \cos (\hat{\omega}_b t) +3 \rho \omega_{q_0}^4 A^2 \cos^3 (\hat{\omega}_b t)=0,
\end{equation}
where $C_{q_0,q_0,q_0,q_0} = \omega_{q_0}^4$, and other terms are neglected.
Applying the trigonometric identity $\cos^3 (\hat{\omega}_b t)=[3 \cos (\hat{\omega}_b t)+\cos (2 \hat{\omega}_b t)]/4$, we obtain
\begin{equation}\label{A15}
\left[\frac{9}{4} \rho \omega_{q_0}^4 A^2-(\hat{\omega}_b^2-\omega_{q_0}^2)\right] \cos (\hat{\omega}_b t) +\frac{3}{4} \rho \omega_{q_0}^4 A^2 \cos (2 \hat{\omega}_b t)=0.
\end{equation}
To eliminate the secular term, the coefficient in front of $\cos (\hat{\omega}_b t)$ must vanish, giving
\begin{equation}\label{A16}
\hat{\omega}_b=\omega_{q_0} \sqrt{1+\frac{9}{4} \rho \omega_{q_0}^2 A^2} =\omega_{q_0} \left[1+\frac{9}{8} \rho \omega_{q_0}^2 A^2+O(\rho^2)\right],
\end{equation}
where $\rho=\beta/[2(2N+1)]$ is a small parameter. Finally, the QB frequency is given by
\begin{equation}\label{A17QB_frequency}
\hat{\omega}_b = \omega_{q_0} \left[ 1 + \frac{9\beta E_{q_0}}{8(2N+1)}  + O(\rho^2) \right],
\end{equation}
where $E_{q_0} = \omega_{q_0}^2 A_{q_0}^2/2$.
Equation (\ref{A17QB_frequency}) is just the Eq. (\ref{QBfrequency}) in the main text, providing the explicit form of the QB frequency.

\section{Stability Analysis of QBs}
\label{AP_C}

This section presents a derivation about the analytical expression for the Floquet multipliers by incorporating the mass difference, following the scheme outlined in Ref. \cite{Flach2006PhysRevE}. Starting from Eq. (\ref{Mathieu}),
\begin{equation}\label{C1}
\ddot{\bm{\xi}} + \bm{\Omega}^2\bm{\xi} + h[1+\cos(2\hat{\omega}_b t)] \bm{B}\bm{\xi} = O(\rho^2, \bm{\xi}),
\end{equation}
we analyze the dynamics governing the perturbation $\bm{\xi}$.
In the limit as $h \to 0$, focusing on the primary resonance, the frequency $\hat{\omega}_b$ can be expressed as
\begin{equation}\label{C2}
\hat{\omega}_b  = \frac{\omega_k + \omega_l}{2}(1 + \delta),
\end{equation}
where the parameter $\delta$ is  of order $O(h)$. We search for solutions of Eq. (\ref{C1}) with the following form
\begin{equation}\label{C3}
    \bm{\xi} = \sum_{m=-\infty}^{\infty} \bm{f}^m e^{ [ z+ {\rm  i}{\omega}_k (1 + \delta)+ {\rm i} m 2\hat{\omega}_b] t} + {\rm c.c.},
\end{equation}
where $\bm{f}^m = (f_q^m)$ are unknown amplitudes, and $z = O(h)$ is assumed to be a small unknown complex number. The closest primary resonance occurs at $k = q_0 - 1$ and $l = q_0 + 1$, leading to the expression \cite{Flach2006PhysRevE}
\begin{equation}
\eqalign{z_{1,2} = &\pm \frac{1}{2} \sqrt{\frac{h^2}{4} \omega_{q_0-1} \omega_{q_0+1} - (h-\delta)^2 (\omega_{q_0-1} + \omega_{q_0+1})^2} \cr & -{\rm i} \frac{(h - \delta)(\omega_{q_0+1} - \omega_{q_0-1})}{2}+O(h^2).}
\label{C4}
\end{equation}
Both \( h \) and \( \hat{\omega}_b \) are dependent on nonlinearity strength \( \beta\varepsilon \), which defines a curve that begins at the point \( (2\omega_{q_0}, 0) \) on the \( (2\hat{\omega}_b, h) \) plane. The intersection of this curve with the resonance band marks the region where QBs become unstable.

For the acoustic branch, the dispersion relation is given by
\begin{equation}\label{C5}
\omega^2_{q_0} = {\frac{2}{1 - \Delta m^2} \left[1 - \sqrt{1 - (1 - \Delta m^2) \sin^2(q_0 \theta )}\right]},
\end{equation}
where $\theta = \pi/(2N+1)$. Expanding $\omega_{q_0}(\theta )$ near $\theta = 0$ using a Taylor series up to the fourth order yields:
\begin{equation}\label{C6}
\omega_{q_0}= q_0 \theta\left(1- \frac{1 + 3\Delta m^2}{24} q_0^2  \theta^2 \right)+O(\theta^5).
\end{equation}
The following approximations are obtained
\begin{equation}\label{C7}
   \eqalign{\omega_{q_0+1} \omega_{q_0-1}  &= q_0^2 \theta^2 \left[1 + O(\theta^2)\right], \cr
   \omega_{q_0+1} + \omega_{q_0-1} &= 2 q_0 \theta -\frac{1+3 \Delta m^2}{12}q_0(3+q_0^2)\theta^3+O(\theta^4).}
\end{equation}
Near the bifurcation point, where the term under the square root in Eq.  (\ref{C4}) approaches zero, we have $h = O(\theta^2)$. Using Eq. (\ref{C2}), $\delta$ can be expressed as
\begin{equation}\label{C10}
\delta = \frac{2\hat{\omega}_b }{\omega_{q_0+1} + \omega_{q_0-1}} - 1 = \frac{3}{4} h + \frac{1 + 3\Delta m^2}{8} \theta^2+O(\theta^4) .
\end{equation}
Substituting Eqs. (\ref{C7}) and (\ref{C10}) into  Eq. (\ref{C4}), we arrive at
\begin{equation}\label{C11}
z_{1,2} = \pm \frac{1}{2} q_0 \theta^3 \frac{1 + 3\Delta m^2}{4} \sqrt{R - 1 + O(\theta^2)} - {\rm i} \,{\rm Im}[z_{1,2}],
\end{equation}
where
\begin{equation}\label{C12}
R = \frac{h}{\theta^2} \frac{4}{(1 + 3\Delta m^2)}.
\end{equation}
$R$ is related to the system's energy $E_{q_0}$ and the nonlinearity parameter $\beta$ as
\begin{equation}\label{C15}
R = \frac{6\beta E_{q_0} (2N+1)}{\pi^2 (1 + 3\Delta m^2)}.
\end{equation}
The absolute value of the Floquet multipliers involved in the resonance is given by
\begin{equation}\label{C13}
|\mu_{\pm}| = \exp\frac{2\pi {\rm Re}[z_{1,2}]}{\hat{\omega}_b}.
\end{equation}
Since $2\pi/\hat{\omega}_b {\rm Re}[z_{1,2}] \to 0$, we have $\mu_{\pm} = 1 + 2\pi/\hat{\omega}_b {\rm Re}[z_{1,2}]$, which ultimately arrives at
\begin{equation}\label{C14}
\mu_{\pm} = 1 \pm \frac{\pi^3 (1 + 3\Delta m^2)}{4(2N+1)^2} {\rm Re}\left[\sqrt{R - 1 + O(\theta^2)}\right].
\end{equation}
This equation is Eq. (\ref{Floquetmultiplier}) in the main text.

\section*{References}
\bibliographystyle{unsrt}
\bibliography{QBreference.bib}

\end{document}